\documentstyle[12pt]{article}
\def\be{\begin{equation}}
\def\ee{\end{equation}}
\def\beq{\begin{eqnarray}}
\def\eeq{\end{eqnarray}}
\def\lsim{\:\raisebox{-0.5ex}{$\stackrel{\textstyle<}{\sim}$}\:}
\def\gsim{\:\raisebox{-0.5ex}{$\stackrel{\textstyle>}{\sim}$}\:}
\begin{document}
\begin{flushright}
{\sl TIFR/TH/95-35}
\end{flushright}
\bigskip\bigskip
\begin{center}
{\Large{\bf Sharpening up the Charged Higgs Boson Signature using Tau
Polarization at LHC}} \\[2cm]
{\large Sreerup Raychaudhuri and D.P. Roy}\footnote{E-mail:
dproy@theory.tifr.res.in}  \\[1cm]
Theoretical Physics Group \\
Tata Institute of Fundamental Research \\
Homi Bhabha Road, Bombay - 400 005, India \\[4cm]
{\large Abstract}
\end{center}
\medskip

The opposite states of the $\tau$ polarization resulting from the
charged Higgs boson and the $W$ boson decays can be exploited to
enhance the $H^\pm$ signal in the inclusive 1-prong hadronic decay
channel of $\tau$.  We suggest practical methods of sharpening up the
$H^\pm$ signature in the top quark decay at LHC using this idea.  As a
result one can carry on the charged Higgs boson search to within $\sim
20 GeV$ of the parent top quark mass over the full parameter space of
the MSSM.

\newpage

\section {INTRODUCTION}
\medskip

A direct top quark signal has been recently observed at the Tevatron
collider [1] with
\be
m_t \simeq 175 \ GeV,
\ee
which also agrees with the indirect estimate of top quark mass from
LEP [2].  One expects a couple of dozens of clearly identifiable top
quark events at the end of the current Tevatron run, which would go up
to a few hundred following the upgradation of its luminosity via the
main injector.  The corresponding number of identifiable top quark
events at LHC is expected to be of the order of a million per year --
i.e. similar to the rate of $Z$ boson events at LEP.  Thus the LHC is
expected to serve as a top quark factory, which will enable us to make
a detailed study of its decay and in particular to search for new
particles in top quark decay.  There has been a good deal of recent
interest in the search for one such new particle, for which the top
quark decay provides by far the best discovery limit -- i.e. the
charged Higgs boson $H^\pm$ of the supersymmetric standard model [3].

There have been several exploratory works on $H^\pm$ search in the top
quark decay at Tevatron and LHC energies [4-7].  They are based on one
of the two distictive properties of $H^\pm$ vis a vis the $W^\pm$
boson -- (i) the preferential $H^\pm$ decay into the $\tau\nu$ channel
relative to $e\nu$ or $\mu\nu$ [4,5], and (ii) the opposite states of
$\tau$ polarization resulting from $H^\pm$ and $W^\pm$ decays [6,7].
In a recent work [8] we have suggested methods of sharpening up the
$H^\pm$ signature in top quark decay by combining these two properties
and applied them to $H^\pm$ search at Tevatron upgrade.  Even with the
best signature, however, the prospect of $H^\pm$ search at Tevatron
upgrade was seen to be limited to a small part of the parameter space.
In the present work we shall analyse the prospect of $H^\pm$ search in
top quark decay at LHC using these ideas.  We shall see below that in
this case one gets a viable signature over practically the full
parameter space of $H^\pm$ coupling.  Moreover with a supplementary
constraint on the accompanying hadronic jets the signature remains
viable over most of the kinematically allowed range of $H^\pm$ mass.

\section{CHARGED HIGGS SIGNAL IN TOP QUARK DECAY}
\medskip

We shall concentrate on the charged Higgs boson of the minimal
supersymmetric standard model (MSSM).  Its
couplings to fermions are given by
\beq
{\cal L} = \frac{g} {\sqrt{2} m_W} H^+ \Bigg\{\cot \beta \ V_{ij}
m_{u_i} \bar u_{i} d_{jL} &+& \tan \beta \ V_{ij} m_{d_j} \bar u_{i} d_{jR}
\nonumber \\[2mm] &+& \tan \beta \ m_{\ell_j} \bar\nu_{j} \ell_{jR}\Bigg\}
+ H.c.
\eeq
where $V_{ij}$ are the Kobayashi-Maskawa (KM) matrix elements and $\tan
\beta$ is the ratio of the vacuum expectation values of the two Higgs
doublets.  The QCD corrections are taken into account in the leading log
approximation by substituting the quark mass parameters by their running
masses evaluated at the $H^\pm$ mass scale [5].  Perturbative
limits on the $tbH$ Yukawa couplings of Eq. (2), along with the
constraints from the low energy processes like $b \rightarrow s\gamma$ and
$B_d-\bar B_d$ mixing, imply the limits [9]
\be
0.4 < \tan\beta < 120.
\ee
In the most predictive form of MSSM, characterised by a common SUSY
breaking mass term at the grand unification point, one gets stronger
limits [10]
\be
1 < \tan\beta < m_t/m_b.
\ee
Such a lower bound also follows from requiring the perturbative limit on
the $tbH$ Yukawa coupling to hold upto the unification point [11].
However, we shall consider the full range (3) of the
phenomenologically allowed $\tan\beta$ parameter.

In the diagonal KM matrix approximation, one gets the decay widths
\beq
\Gamma_{t \rightarrow bW} &=& \frac{g^2}{64\pi m^2_W m_t}
\lambda^{\frac{1}{2}}
\left(1,\frac{m^2_b}{m^2_t},\frac{m^2_W}{m^2_t}\right)\nonumber
\\[2mm] & &
\left[m^2_W (m^2_t + m^2_b) + (m^2_t - m^2_b)^2 - 2m^4_W\right] \\
\Gamma_{t\rightarrow bH} &=& \frac{g^2}{64\pi m^2_W m_t}
\lambda^{\frac{1}{2}}
\left(1,\frac{m^2_b}{m^2_t},\frac{m^2_H}{m^2_t}\right) \nonumber \\
[2mm] & & \left[(m^2_t \cot^2\beta + m^2_b \tan^2\beta) (m^2_t + m^2_b -
m^2_H) - 4m^2_t m^2_b\right] \\
\Gamma_{H \rightarrow \tau\nu} &=& \frac{g^2m_H}{32\pi m^2_W} m^2_\tau
\tan^2 \beta \\
\Gamma_{H \rightarrow c\bar s} &=& \frac{3g^2 m_H}{32\pi m^2_W}
\left(m^2_c \cot^2 \beta + m^2_s \tan^2 \beta\right).
\eeq
\noindent From these one can construct the relevant branching fractions
\be
B_{t \rightarrow bH} = \Gamma_{t \rightarrow bH}\big/\left(\Gamma_{t
\rightarrow bH} + \Gamma_{t \rightarrow bW}\right)
\ee
\be
B_{H \rightarrow \tau \nu} = \Gamma_{H \rightarrow
\tau\nu}\big/\left(\Gamma_{H \rightarrow \tau\nu} + \Gamma_{H
\rightarrow c\bar s}\right).
\ee
It is the product of these two branching fractions that controls the size
of the observable charged Higgs signal.  The $t \rightarrow bH$ branching
fraction has a pronounced dip at
\be
\tan\beta = (m_t/m_b)^{\frac{1}{2}} \simeq 6,
\ee
where (6) has a minimum.  Although this is partly compensated by a large
value of the $H \rightarrow \tau\nu$ branching fraction, which is $\simeq
1$ for $\tan\beta > 2$, the product still has a significant dip at (11).
Consequently the predicted charged Higgs signal will be very weak around
this point as we shall see below.

The basic process of interest is $t\bar t$ pair production through
gluon-gluon (or quark-antiquark) fusion followed by their decay into
charged Higgs or $W$ boson channels, i.e.
\be
gg \rightarrow t\bar t \rightarrow b\bar b (H^+H^-, H^\pm W^\mp, W^+ W^-).
\ee
The $\tau$ decay of one or both the charged bosons leads to a single
$\tau$, $\tau\tau$ or $\ell\tau$ final state, where $\ell$ denotes $e$ and
$\mu$.  Each of these final states is accompanied by several hadronic
jets and a large missing-$E_T$ (transverse energy) due to the neutrinos.

A brief discussion of the $\tau$-identification at hadron colliders is in
order here. Starting with a missing-$E_T$ trigger, the UA1, UA2 and CDF
experiments have been able to identify $\tau$ as a narrow jet in its
hadronic decay mode [12,13].  In particular the CDF experiment has
used the narrow jet cut to reduce the QCD jet background by an order of
magnitude while retaining most of the hadronic $\tau$ events.  Moreover,
since the hadronic $\tau$ and QCD jet events dominantly populate the
1-prong and multi-prong channels respectively, the prong distribution of
the narrow jets can be used to distinguish the two.  In particular
restricting to 1-prong jets reduces the QCD background by another
order of magnitude with very little loss to the hadronic $\tau$
signal.  In this way the CDF group [13] has been able to reduce the
QCD background to a few tens of events in a data sample of integrated
luminosity $\sim 4 pb^{-1}$, which could be subtracted out by
extrapolation from higher prong channels.  Consequently they were able
to identify the $W \rightarrow \tau\nu$ events and test $W$
universality as well as put some modest constraints on top and $H^\pm$
masses from the level of the residual $\tau\nu$ events.  In the
present case, however, one would be looking for a few tens of hadronic
$\tau$ events in a data sample of over 1000 times higher integrated
luminosity and 10 times higher QCD jet cross-section.  So the QCD jet
background cannot be controlled by the above method, even with $b$
identification.  Therefore one
cannot use the single $\tau$ channel for the charged Higgs search and even
the $\tau\tau$ channel can be at best marginal.  The best charged Higgs
signature is provided by the $\ell\tau$ channel.  The largest background
comes from $W \rightarrow \ell\nu$ accompanied by QCD jets, which can be
easily controlled by the above mentioned jet angle and multiplicity cuts.
Besides the hard isolated lepton $\ell$ provides a more robust trigger
than the missing-$E_T$, particularly at the LHC.  Therefore we shall
concentrate mainly on the $\ell\tau$ channel.

The $\ell\tau$ and $\tau\tau$ channels correspond to the leptonic decay of
both the charged bosons in (12), i.e.
\beq
& H^+ & \ H^- \ \ , \ \ H^+ \ \ \ W^- \ \ , \ \ H^- \ \ \ W^+ \ \ ,
\ \ \ W^+ \ \ \ \ \ \ \ W^- ,\nonumber \\[2mm]
& \downarrow & \ \ \downarrow \ \ \ \ \ \ \ \
\downarrow \ \ \ \ \ \ \downarrow \ \ \ \ \ \ \ \ \
\downarrow \ \ \ \ \ \ \downarrow \ \ \ \ \ \ \ \ \ \ \
\downarrow \ \ \ \ \ \ \ \ \ \ \downarrow \nonumber \\[2mm]
& \tau^+_L & \ \tau^-_R \ \ \ \ \ \tau^+_L \ \ \ \tau^-_L,\ell^- \ \ \ \ \
\tau^-_R \ \ \ \tau^+_R,\ell^+ \ \ \ \ \ \tau^+_R,\ell^+ \ \ \
\tau^-_L,\ell^-
\eeq
where $L$ and $R$ stand for left and right handed helicities of $\tau$.
By convention,
\be
P_\tau \equiv P_{\tau^-} = -P_{\tau^+}, \ \ \ P_{\tau^\pm} =
\frac{\sigma_{\tau^\pm_R} - \sigma_{\tau^\pm_L}}{\sigma_{\tau^\pm_R} +
\sigma_{\tau^\pm_L}}.
\ee
For the $\ell\tau$ channel of our interest the signal and the background
come from the $HW$ and $WW$ terms respectively.  They correspond to
exactly opposite states of $\tau$ polarization, {\em i.e.}
\be
P^H_\tau = + 1, \ \ \ P^W_\tau = -1.
\ee
Consequently the use of the tau polarization effect for enhancing the
signal to background ratio is particularly simple in this case as we shall
see below.  It may be noted here that the $\tau\tau$ channel has a better
signal to background ratio because of the $HH$ contribution as well as the
enhancement of $WH$ relative to $WW$ by a combinatorial factor of 2 [5].  On
the other hand the polarization distinction is less simple. While both the
$\tau$'s in the background have negative polarization one or both of them
have positive polarization in the signal.  Nonetheless the method of
enhancing the signal to background ratio by the $\tau$ polarization effect
discussed below can be extended to this channel, provided one can identify
the $\tau\tau$ events from the QCD background.
\bigskip

\section{TAU POLARIZATION EFFECT}
\medskip

We shall concentrate on the 1-prong hadronic decay channel of $\tau$,
which is best suited for $\tau$ identification.  It accounts for 80\% of
hadronic $\tau$ decays and 50\% of overall $\tau$ decays.  The main
contributors to the 1-prong hadronic $\tau$ decay are [2]
\beq
\tau^\pm & \rightarrow & \pi^\pm \nu_\tau  \ (12.5\%)  \\
\tau^\pm & \rightarrow & \rho^\pm \nu_\tau \rightarrow \pi^\pm \pi^0
\nu_\tau  \ (24\%)  \\
\tau^\pm & \rightarrow & a^\pm_1 \nu \rightarrow \pi^\pm \pi^0 \pi^0 \nu_\tau
 \ (7.5\%)
\eeq
where the branching fractions for the $\pi$ and $\rho$ channels include
the small contributions from the $K$ and $K^\star$ channels respectively,
since they have identical polarization effects.  Note that only half the
$a_1$ decay channel contributes to the 1-prong configuration.  The masses
and widths of $\rho$ and $a_1$ are [2]
\be
m_\rho (\Gamma_\rho) = 770 (150) \ {\rm MeV}, \ \ \ m_{a_1} (\Gamma_{a_1})
= 1260 (400) \ {\rm MeV}.
\ee
One sees that the three decay processes (16,17,18) account for about 90\%
of the 1-prong hadronic decay of $\tau$.  Thus the inclusion of $\tau$
polarization effect in these processes will account for its effect in the
inclusive 1-prong hadronic decay channel to a good approximation.

The formalism relating $\tau$ polarization to the momentum distribution of
its decay particles in (16,17,18) has been widely discussed in the
literature [6,7,8,14,15].  We shall only discuss the
main formulae relevant for our analysis.  A more detailed account can be
found in a recent paper by Bullock, Hagiwara and Martin [7],
which we shall closely follow.  For $\tau$ decay into $\pi$ or a vector
meson $(\rho,a_1)$, one has
\beq
\frac{1}{\Gamma_\pi} \frac{d\Gamma_\pi} {d \cos\theta}
&=& \frac{1}{2} (1 + P_\tau \cos \theta)
\\
\frac{1}{\Gamma_v} \frac{d\Gamma_{vL}}{d\cos\theta}
&=& \frac{\frac{1}{2} m^2_\tau}{m^2_\tau + 2m^2_v} (1 + P_\tau \cos\theta)
\\
\frac{1}{\Gamma_v} \frac{d\Gamma_{vT}}{d \cos\theta}
&=& \frac{m^2_v}{m^2_\tau + 2m^2_v} (1 - P_\tau \cos\theta)
\eeq
where $v$ stands for the vector meson and $L,T$ denote its longitudinal
and transverse polarization states.  The angle $\theta$ measures the
direction of the meson in the $\tau$ rest frame relative to the $\tau$
line of flight, which defines its polarization axis.  It is related to the
fraction $x$ of the $\tau$ energy-momentum carried by the meson in the
laboratory frame, {\it i.e.}
\be
\cos\theta = \frac{2x - 1 - m^2_{\pi,v}/m^2_\tau}{1 - m^2_{\pi,v}/m^2_\tau}.
\ee
Here we have made the collinear approximation $m_\tau \ll p_\tau$, where
all the decay products emerge along the $\tau$ line of flight in the
laboratory frame.

The above distribution (20-22) can be simply understood in terms of
angular momentum conservation.  For $\tau^-_{R(L)} \rightarrow \nu_L \
\pi^-$, $v^-_{\lambda=0}$ it favours forward (backward) emission of $\pi$
or longitudinal vector meson, while it is the other way round for
transverse vector meson emission $\tau^-_{R(L)} \rightarrow \nu_L
v^-_{\lambda=-1}$.  Thus the $\pi^\pm$s coming from $H^\pm$ and $W^\pm$
decays peak at $x=1$ and $0$ respectively and $\langle x_\pi\rangle_H =
2\langle x_\pi\rangle_W = 2/3$.  Although the clear separation between the
signal and the background peaks disappears after convolution with the
$\tau$ momentum, the relative size of the average $\pi$ momenta remains
unaffected, {\em i.e.}
\be
\langle p^T_\pi \rangle_H \simeq 2 \langle p^T_\pi \rangle_W \ \ {\rm
for} \ \ m_H \simeq m_W.
\ee
Thus the $\tau$ polarization effect (20) is reflected in a significantly
harder $\pi^\pm$ momentum distribution for the charged Higgs signal
compared to the $W$ boson background.  The same is true for the
longitudinal vector mesons; but the presence of the transverse component
dilutes the polarization effect in the vector meson momentum distribution
by a factor (see eqs. 21,22)
\be
\frac{m^2_\tau - 2m^2_v}{m^2_\tau + 2m^2_v}.
\ee
Consequently the effect of $\tau$ polarization is reduced by a factor of
$\sim 1/2$ in $\rho$ momentum distribution and practically washed out in
the case of $a_1$.  Thus the inclusive 1-prong $\tau$ jet resulting from
(16-18) is expected to be harder for the $H^\pm$ signal compared to the
$W$ boson background; but the presence of the transverse $\rho$ and $a_1$
contributions makes the size of this difference rather modest.  We shall
see below that it is possible to suppress the transverse $\rho$ and $a_1$
contributions and enhance the difference between the signal and the
background in the 1-prong hadronic $\tau$ channel even without identifying
the individual mesonic contributions to this channel.

The key feature of vector meson decay, relevant for the above purpose, is
the correlation between its state of polarization and the energy sharing
among the decay pions.  The transverse $\rho$ and $a_1$ decays favour
even sharing of energy among the decay pions, while the longitudinal
$\rho$ and $a_1$ decays favour asymmetric configurations where the
charged pion carries either very little or most of the vector meson
energy.  It is easy to derive this quantitatively for the $\rho$
decay.  But $a_1$ decay is more involved.  One can show from general
considerations that the $a_{1T(L)} \rightarrow 3\pi$ decay favours the
plane of the 3 pions in the $a_1$ rest frame being normal to
(coincident with) the $a_1$ line of flight [15].  This agrees
qualitatively with the above pattern of energy sharing.  But one has
to assume a dynamical model for $a_1$ decay to get a more quantitative
result.  We shall follow the model of Kuhn and Santamaria, based on
the chiral limit (conserved axial-vector current approximation), which
provides a good description of the $a_1 \rightarrow 3\pi$ data [16].
One gets very similar pion energy distributions from the alternative
model of Isgur et al [17], as shown in [7].  A detailed account of the
$\rho$ and $a_1$ decay formalisms can be found in [7,8] along with the
prescriptions for incorporating the finite $\rho$ and $a_1$ widths.
We shall only summarise the results below.

Fig. 1 shows the $\rho$ and $a_1$ decay distributions in the
energy-momentum fraction $x'$, carried by the charged pion.  The
distributions are shown for both longitudinal and transverse
polarization states of the vector mesons.  The transverse $\rho$ and
$a_1$ distributions are clearly seen to vanish at the end points and
peak in the middle, reflecting equipartition of energy-momentum among
the decay pions.  In contrast the longitudinal $\rho$ distribution
shows pronounced peaks near the end points $x' = 0$ and $1$, and the
longitudinal $a_1$ at the former point.  Note also that the direct
pionic decay of $\tau$ (16) can be formally looked upon as a delta
function contribution at $x' = 1$ in this figure.  Thus one can
suppress the unwanted $\rho_T$ and $a_{1T}$ contributions while
retaining the $\pi$ and at least good fractions of $\rho_L$ and
$a_{1L}$ by restricting to the regions $x' \simeq 0$ and $1$.  We
shall see below how this can be achieved even without identifying the
individual mesonic contributions in $\tau$ decay.
\bigskip

\section{STRATEGY, RESULTS AND DISCUSSION}
\medskip

As mentioned earlier, we are interested in the inclusive 1-prong
hadronic decay of $\tau$, which is dominated by the $\pi^\pm$,
$\rho^\pm$ and $a^\pm_1$ contributions (16,17,18).  It results in a
thin 1-prong hadronic jet ($\tau$-jet) consisting of a charged pion
along with $0,1$ or $2$ $\pi^0$'s respectively.  Since all the pions
emerge in a collinear configuration, one can neither measure their
invariant mass nor the number of $\pi^0$'s.  Thus it is not possible
to identify the three mesonic states.  But it is possible to measure
the energy of the charged track and the accompanying neutral energy
separately, either by measuring the momentum of the former in the
tracking chamber and the total energy deposit in the electromagnetic
and hadronic calorimeters serrounding it or from the showering
profiles in these electromagnetic and hadronic calorimeters [18].
Thus one has to devise a strategy to suppress the transverse vector
meson contributions using these two pieces of information.  We shall
consider two such strategies below.  In either case a rapidity and a
transverse energy cut of
\be
|\eta| \ < \ 3 \ \ \ {\rm and} \ \ \ E_T \ > \ 20 \ GeV
\ee
will be applied on the $\tau$-jet as well as the tagging lepton
$\ell$, where $E_T$ includes the neutral contribution to the former
[18].  We shall also apply isolation cuts to ensure that there are no
hadronic jets within a cone of radius $\Delta R = (\Delta \eta^2 +
\Delta \phi^2)^{1/2} = 0.4$ around the $\tau$-jet and the tagging
lepton.

The first strategy is to impose a calorimetric isolation cut on the
$\tau$-jet, which requires the neutral $E_T$ accompanying the charged
track within a cone of $\Delta R = 0.2$ to be less than $5
GeV$,\footnote{Depending on the energy resolution of the calorimeters,
this can be increased upto $10 GeV$ without affecting the results
significantly.} i.e.
\be
E_T^{ac} \equiv E^0_T \ < \ 5 \ GeV.
\ee
As we see from Fig. 1, this cut eliminates the $\rho_T$ and $a_{1T}$
contributions along with the $x' \simeq 0$ peaks of $\rho_L$ and
$a_{1L}$.  It retains only the $\pi$ and the $x' \simeq 1$ peak of the
$\rho_L$ contribution.  This results in a substantially harder signal
cross-section relative to the background, but at the cost of a factor
of $\sim 2$ drop in the signal size.

The second strategy is to plot the $\tau$-jet events satisfying (26)
as a function of
\be
\Delta E_T = |E_T^{ch} - E^0_T|,
\ee
i.e. the difference between the $E_T$ of the charged track and the
accompanying neutral $E_T$ instead of their sum.  It is clear from
Fig. 1 that the even sharing of the transverse $\rho$ and $a_1$
energies among the decay pions imply a significantly softer $\Delta
E_T$ distribution for $\rho_T$ and $a_{1T}$ relative to $\rho_L$ and
$a_{1L}$.  This results in a substantially harder signal cross-section
relative to the background when plotted against $\Delta E_T$ instead
of $E_T$.  Moreover this is achieved at no cost to the signal size
unlike the previous case.

In comparing the two methods one notes that the first is easier to
implement and besides it helps to suppress the level of QCD jet
background as well.  On the other hand the second method has the
advantage of a factor of $\sim 2$ larger cross-section.  While
studying the $H^\pm$ signature at the Tevatron upgrade in [8], we had
found the second method more viable in view of the limited size of the
$t\bar t$ signal there.  Since the size of this signal will be very
large at the LHC, however, both the methods will be equally viable as
we shall see below.

We have estimated $H^\pm$ signal and the $W^\pm$ background
cross-sections at the LHC energy of
\be
\sqrt{s} = 14 \ TeV
\ee
using a parton level Monte Carlo program with the recent structure
functions of [19].  In stead of the differential cross-section in
$E_T$ (or $\Delta E_T$), we have plotted the corresponding integrated
cross-sections
\be
\sigma (E_T) = \int^\infty_{E_T} {d\sigma \over dE_T} dE_T
\ee
against the cut-off value of $E_T$ (or $\Delta E_T$).  Fig. 2 shows
these cross-sections for
\be
\tan\beta = 3 \ \ \ {\rm and} \ \ \ m_H = 120, 140 \ GeV
\ee
(a) for the raw signal, (b) using the calorimetric isolation cut (27)
and (c) using $\Delta E_T$ instead of $E_T$.

There is a clear hardening of the signal curves relative to the
background as one goes from Fig. 2a to 2b or 2c.  One could of course
see a similar hardening in the corresponding differential
cross-sections.  But the present plots are better suited to compare
the relative merits of the three methods in extracting the signal from
the background.  For this purpose the cut-off value of $E_T$ $(\Delta
E_T)$ is to be so chosen that one gets a viable signal to background
ratio, i.e.
\be
H^\pm \ {\rm signal}/W^\pm \ {\rm background} \ \geq 1.
\ee
The resulting signal cross-section, as given by the cross-over point
between the signal and background curves, is a reasonable criterion
for the merit of the method.  One clearly sees that this point is
reached at a much larger value of the cut-off in Fig. 2a,
corresponding to a far greater sacrifice to the signal size, than in
2b or 2c.  The resulting signal cross-sections for $m_H = 120$ and
$140 \ GeV$ are $\sim 20 \ fb$ and less than $1 \ fb$ in Fig. 2a,
going up to $\sim 300$ and $50 \ fb$ respectively in Fig. 2b and about
double these values in Fig. 2c.  It is remarkable that a simple
calorimetric isolation cut (27) can enhance the signal-background
separation so much and increase the effective signal cross-section by
over an order of magnitude.  Of course the signal cross-section
obtained via the $\Delta E_T$ distribution is still larger by a factor
of $\sim 2$, as anticipated earlier.

To probe the $H^\pm$ discovery limits at LHC using the three methods,
we have estimted the corresponding signal cross-sections, satisfying (32), as
functions of $\tan\beta$.  These are shown in Fig. 3a,b,c for a set of
$H^\pm$ masses,
\be
m_H = 80,100,120,140,150,160 \ GeV.
\ee
There is a clear dip at $\tan\beta \simeq 6$ as anticipated in (11).
One sees a gap in the $\tan\beta$ space around this region where the raw
signal of Fig. 3a is clearly not viable.  But the improved signals
obtained via the calorimetric isolation cut (Fig. 3b) or the $\Delta
E_T$ distribution (Fig. 3c) remain $\gsim 100 (1) \ fb$ for $m_H = 120
(140) \ GeV$ throughout the $\tan\beta$ space.  It may be noted here
that one expects an integrated luminosity of
\be
\int L \ dt \sim 100 \ fb^{-1}
\ee
from the high luminosity run of the LHC.  For a signal size of $\sim 1
\ fb$ satisfying (32), it corresponds to $H^\pm$ signal and $W^\pm$
background events of $\sim 100$ each.  Since the latter can be
predicted from the number of $t\bar t$ events in the dilepton $(\ell^+
\ell^-)$ channel using $W$ universality, this will correspond to a
$\sim 10 \sigma$ signal for the $H^\pm$ boson.  Thus a signal size of
$\sim 1 \ fb$ in Fig. 3 will constitute a viable signal for the high
luminosity run of LHC.  This means that the improved signatures for
$H^\pm$ boson search at LHC are viable upto $m_H = 140 \ GeV$ over the
full $\tan\beta$ space.

It may appear from Fig. 3 that for extreme values of $\tan \beta$
($\lsim 1$ or $\gsim 50$), where the raw signal of 3a is already
large, there is no advantage in going to 3b or 3c.  It should be noted
however that in this case the $\tau$ polarization effect can serve as
an independent test for the $H^\pm$ signal.  The hardening (softening)
of the signal (background) cross-section of Fig. 2 or equivalently the
corresponding differential cross-section [8], as one imposes the
calorimetric isolation cut or goes to the $\Delta E_T$ variable, is a
distinctive prediction of the $\tau$ polarization effect that holds
independent of $\tan\beta$.

Finally, one can push the viability of these two signatures to still
higher values of $m_H$ with a suitable cut on the accompanying
hadronic jets.  For this purpose one exploits the fact that for $m_H
\simeq m_t$ the accompanying $b$-jet in the $t \rightarrow bH(W)
\rightarrow b\tau\nu$ decay is necessarily soft for the $H$ signal but
not the $W$ background [5].  Thus the $WW$ background can be
suppressed without sacrificing the $HW$ signal by imposing a kinematic
cut of
\be
E_T^{jet} \ < \ 30 \ GeV
\ee
on all but one of the accompanying hadronic jets.  Of course in the
process one would be sacrificing both the signal and background events
which are accompanied by a hard QCD jet, which implies a reduction of
the signal size without affecting the signal to background ratio.  But
we do not expect this reduction factor to be very large.  Moreover if
there is reasonable $b$ identification efficiency at the LHC, then one
can by pass this problem by imposing this cut on one of the identified
$b$-jets.

Fig. 4a,b,c shows the $\tan\beta$ distribution of the signal
cross-sections satisfying (32), after this kinematic cut.  The
cross-sections are shown only for the high values of $m_H$ ($140,150$
and $160 \ GeV$), for which the cut is relevant.  There is again a
large gap in the $\tan\beta$ space where the raw signal is not viable
(Fig. 4a).  But the improved signals via the calorimetric isolation
cut (Fig. 4b) or the $\Delta E_T$ distribution (Fig. 4c) remain $> \
10(1) \ fb$ for $m_H = 140 (150) \ GeV$ throughout the $\tan\beta$
parameter space.  Thus they provide unambiguous signatures for $H^\pm$
boson search upto $m_H = 140$ and $150 \ GeV$ at the low and high
luminosity runs of LHC, corresponding to integrated luminosities of
$10$ and $100 \ fb^{-1}$ respectively.   In fact for the high
luminosity run the signatures remain viable over the full $\tan\beta$
space upto a $H^\pm$ mass of $155 \ GeV$ -- i.e. within $20 \ GeV$ of
the parent top quark mass.

In summary, we have explored the prospect of charged Higgs boson
search in top quark decay at the LHC, taking advantage of the opposite
states of $\tau$ polarization resulting from the $H^\pm$ and $W^\pm$
decays.  We have concentrated on the most promising channel for
$H^\pm$ search -- i.e. the $\ell \tau$ channel, followed by the
inclusive decay of $\tau$ into a 1-prong hadronic jet.  Two practical
methods of sharpening up the $H^\pm$ signature, using the $\tau$
polarization effect, have been studied.  The resulting signatures are
shown to be viable over the full parameter space of $\tan\beta$ upto
$m_H = 140 \ GeV$.  Moreover with a kinematic cut on the
accompanying hadronic jets, one can stretch their viability upto
$m_H = 150 - 155 \ GeV$ -- i.e. within $20 \ GeV$ of the parent
top quark mass.
\medskip

The work of SR is partially supported by a project (DO
No. SR/SY/P-08/92) of the Department of Science and Technology,
Government of India.

\newpage

\noindent {\large{\bf References}}
\bigskip

\begin{enumerate}
\item[{[~1]}] CDF collaboration: F. Abe et al., Phys. Rev. Lett. 74
(1995) 2626; D$O\!\!\!\!/$ collaboration: S. Abachi et al.,
Phys. Rev. Lett. 74 (1995) 2632.
\item[{[~2]}] Review of Particles Properties, Phys. Rev. D50 (1994)
1173-1826.
\item[{[~3]}] J.F. Gunion, H.E. Haber, G. Kane and S. Dawson, The Higgs
Hunters' Guide, Addison Wesley, Reading, MA (1990).
\item[{[~4]}] V. Barger and R.J.N. Phillips, Phys. Rev. D41 (1990) 884;
A.C. Bawa, C.S. Kim and A.D. Martin, Z. Phys. C47 (1990) 75;
R.M. Godbole and D.P. Roy, Phys. Rev. D43 (1991) 3640.
\item[{[~5]}] M. Drees and D.P. Roy, Phys. Lett. B269 (1991) 155;
D.P. Roy, Phys. Lett. B283 (1992) 403.
\item[{[~6]}] R.M. Barnett et al., Proc. DPF Summer Study on High
Energy Physics, Snowmass (1990); B.K. Bullock, K. Hagiwara and
A.D. Martin, Phys. Rev. Lett. 67 (1991) 3055; D.P. Roy,
Phys. Lett. B277 (1992) 183.
\item[{[~7]}] B.K. Bullock, K. Hagiwara and A.D. Martin,
Nucl. Phys. B395 (1993) 499.
\item[{[~8]}] Sreerup Raychaudhuri and D.P. Roy, Phys. Rev. D (in
press).
\item[{[~9]}] V. Barger, J.L. Hewett and R.J.N. Phillips,
Phys. Rev. D41 (1990) 3421; J.F. Gunion and B. Grzadkowski,
Phys. Lett. B243 (1990) 301; A.J. Buras et al., Nucl. Phys. B337
(1990) 284.
\item[{[10]}] G. Ridolfi, G.G. Ross and F. Zwirner, Proc. ECFA-LHC
Workshop, CERN 90-10, Vol. II, (1990) p. 608.
\item[{[11]}] J. Bagger, S. Dimopoulos and E. Masso,
Phys. Rev. Lett. 55 (1985) 920.
\item[{[12]}] UA2 Collaboration: J. Alitti et al., Phys. Lett. B280
(1992) 137; UA1 Collaboration: C. Albajar et al., Phys. Lett. B251
(1991) 459.
\item[{[13]}] CDF Collaboration : F. Abe et al., Phys. Rev. Lett. 72
(1994) 1977.
\item[{[14]}] Y.S. Tsai, Phys. Rev. D4 (1971) 2821; P. Aurenche and
R. Kinnunen, Z. Phys. C28 (1985) 261; K. Hagiwara, A.D. Martin and
D. Zeppenfeld, Phys. Lett. B235 (1990) 198.
\item[{[15]}] A. Rouge, Z. Phys. C48 (1990) 75.
\item[{[16]}] J.H. Kuhn and A. Santamaria, Z. Phys. C48 (1990) 445.
\item[{[17]}] N. Isgur, C. Morningstar and C. Reader, Phys. Rev. D39
(1989) 1357.
\item[{[18]}] See e.g. CMS Technical Proposal, CERN/LHCC 94-38 (15
Dec. 1994).
\item[{[19]}] A.D. Martin, R.G. Roberts and W.J. Stirling,
Phys. Lett. B306 (1993) 145 and B309 (1993) 492.
\end{enumerate}

\newpage

\begin{center}
{\large\underbar{Figure Captions}}
\end{center}
\bigskip\bigskip

\begin{enumerate}
\item[{Fig. 1.}] Distributions of the $\rho^\pm \rightarrow \pi^\pm
\pi^0$ and $a^\pm_1 \rightarrow \pi^\pm \pi^0 \pi^0$ decay widths in
the energy fraction carried by the charged pion, shown separately for
the transverse and longitudinal states of $\rho$ and $a_1$
polarization.
\item[{Fig. 2.}] The integrated 1-prong hadronic $\tau$-jet
cross-sections are plotted against cut-off values of the jet $E_T$ in
(a) without and (b) with the isolation cut.  They are plotted against
the cutoff-value of $\Delta E_T$ of the jet in (c).  The $H^\pm$
signal ($W^\pm$ background) contributions are shown as solid (dashed)
lines for $m_H = 120 \ GeV$ and dot-dashed (dotted) lines for $m_H =
140 \ GeV$.  We take $\sqrt{s} = 14 \ TeV$ and $\tan\beta = 3$.
\item[{Fig. 3.}] The signal cross-sections of Fig. 2(a,b,c) satisfying
a signal to background ratio $\geq 1$ are shown as functions of
$\tan\beta$ for $m_H = 80,100,120,140,150,160$ $GeV$ by solid, dashed,
dot-dashed, double dot-dashed, dotted and multidot-dashed lines
respectively.
\item[{Fig. 4.}] The signal cross-sections are shown as in Fig. 3, but
with an additional cut of $E_T^{jet} \ < \ 30 \ GeV$ on the second
hardest accompanying jet, for $m_H = 140,150,160 \ GeV$ by solid,
dashed and dot-dashed lines respectively.
\end{enumerate}
\end{document}